\documentclass[12pt,twoside,english]{article}
\usepackage[T1]{fontenc}
\usepackage[latin1]{inputenc}
\usepackage{color}
\usepackage{amsbsy}
\usepackage{amssymb}
\usepackage{babel}
\begin{document}

\title{Electromagnetic-radiation effect on alpha decay}

\author{{\normalsize{}M. Apostol }\\
{\normalsize{}Department of Theoretical Physics, Institute of Atomic
Physics, }\\
{\normalsize{}Magurele-Bucharest MG-6, POBox MG-35, Romania }\\
{\normalsize{}email: apoma@theory.nipne.ro}}

\date{{J. Math. Theor. Phys. 1 155 (2018)}}

\maketitle
\relax
\begin{abstract}
The effect of the electromagnetic radiation on the \textcolor{black}{spontaneous}
charge emission from heavy atomic nuclei is estimated in a model which
may be relevant for proton emission and alpha-particle decay in laser
fields. Arguments are given that the electronic cloud in heavy atoms
screens appreciably the electric field acting on the nucleus and\textcolor{black}{{}
the nucleus \textquotedbl{}sees\textquotedbl{} rather low fields.
In these conditions,} it is shown that the electromagnetic radiation
brings second-order corrections in the electric field to the disintegration
rate, with a slight anisotropy. These corrections give a small enhancement
of the disintegration rate. The case of a static electric field is
also discussed. 
\end{abstract}
\relax



\noindent In the context of an active topical research in laser-related
physics,\cite{key-1}-\cite{key-5} the problem of charge emission
from bound-states under the action of the electromagnetic radiation
is receiving an increasing interest. Some investigations focus especially
on the effect the optical-laser radiation may have on the \textcolor{black}{spontaneous}
alpha-particle decay of the atomic nuclei,\cite{key-6}-\cite{key-11}
or nuclear proton emission,\cite{key-12,key-13} but the area may
be extended to atom ionization or molecular or atomic clusters fragmentation.\cite{key-14}-\cite{key-17}
The aim of the present paper is to estimate the effect of the adiabatically-applied
electromagnetic radiation upon the rate of \textcolor{black}{spontaneous
}nuclear alpha decay and proton emission. Specifically, the paper
is motivated by the interest in computing the rate of tunneling through
a Coulomb potential barrier in the presence of electric fields. It
is claimed that the rate of alpha decay is practically not affected
by electric fields,\cite{key-8} or it is greatly enhanced by strong
electric fields.\cite{key-18} On the other side, the atomic electron
cloud may screen appreciably the external electric fields, such that
the atomic nucleus may experience, in fact, rather low electric fields.
It is this point, related to low electric fields, which may raise
technical difficulties in estimating the small effect of these external
fields upon the alpha decay. 

We adopt a nuclear model with $Z$ protons and $A-Z$ neutrons, where
$A$ is the mass number of the nucleus, moving in the nuclear mean
field. The experiments proceed usually by placing a collection of
heavy atoms in the focal region of a laser beam, and focusing radiation
pulses upon that collection of atoms. We consider an optical-laser
radiation with a typical frequency $\omega$ of the order $10^{15}s^{-1}$
(corresponding to a period $T\simeq10^{-15}s$ and a wavelength $\lambda\simeq0.8\mu m$).
We assume that the motion of the charges under the action of the electromagnetic
radiation remains non-relativistic, \emph{i.e.} $qA_{0}/mc^{2}\ll1$,
where $q$ is the particle charge, $m$ is the particle mass and $A_{0}$
is the amplitude of the vector potential ($c$ denotes the speed of
light in vacuum). For protons in atomic nuclei ($q=4.8\times10^{-10}esu$,
$m\simeq2\times10^{-24}g$, $c=3\times10^{10}cm/s$) this condition
yields a very high electric field $E_{0}=3\times10^{13}V/cm$ ($10^{11}$
electrostatic units), which corresponds to a maximum intensity of
the laser beam in the focal region of the order $I=cE_{0}^{2}/8\pi=10^{24}w/cm^{2}$.
Typically, the duration of the laser pulse is of the order of tens
of radiation period (or longer), such that we may consider the action
of the electromagnetic radiation much longer than the period of the
radiation. The repetition rate of the laser pulses is usually much
longer than the pulse duration. For simplification we consider linearly-polarized
radiation plane waves; the calculations can be extended to a general
polarization. The laser-beam shape or multi-mode operation have little
relevance upon the results presented here.

The electric fields are appreciably screened by the electronic cloud
of the heavy atoms. The screening effects on the thermonuclear reactions,
alpha decay and lifetimes have been considered previously.\cite{key-10,key-11,key-19,key-20}
A convenient means of treating the electron cloud in heavy atoms is
the linearized Thomas-Fermi model.\cite{key-21} According to this
model, the radial electron distribution is concentrated at distance
$R=a_{H}/Z^{1/3}$ (screening distance), where $a_{H}=\hbar^{2}/mq^{2}$
is the Bohr radius and $Z$ is the atomic number ($Z\gg1$); $q$
and $m$ denote the electron charge and mass, respectively. The atomic
binding energy depends on $R$, and the atom exhibits an eigenmode
related to the change in $R$ (a breathing-type mode), with an eigenfrequency
$\omega_{0}\simeq Z\mid q\mid/\sqrt{ma_{H}^{3}}\simeq4.5Z\times10^{16}s^{-1}$
($\hbar\omega_{0}\simeq28Z(eV)$). We recognize in $\omega_{0}$ the
plasma frequency $\simeq4\pi\overline{n}q^{2}/m$ of a mean electron
density $\overline{n}\simeq Z/R^{3}=Z^{2}/a_{H}^{3}$. It corresponds
to the atomic giant-dipole oscillations discussed in Ref. \cite{key-21}.
In the presence of an external electric field $E$ oriented along
the $z$-direction the electrons are displaced by $u$ (with fixed
nucleus), \textcolor{black}{a displacement}\textcolor{red}{{} }which
produces an energy change $\sim z^{2}u^{2}/R^{2}$. By integrating
over $z$, we get a factor $1/\sqrt{3}$ in the eigenfrequency $\omega_{0}$,
as expected. It follows that the displacement $u$ obeys the equation
of motion $\ddot{u}+\Omega^{2}u=qE/m$, where $\Omega=\omega_{0}/\sqrt{3}$;
the internal field is $E_{i}=-4\pi\overline{n}qu$ (polarization $P=\overline{n}qu$
and the dipole moment $p=Zqu$). For $E=E_{0}\sin\omega t$ the solution
of this equation is $u=u_{0}\sin\omega t$, $u_{0}=-(qE_{0}/m)/(\omega^{2}-\Omega^{2}$),
and the internal field is $E_{i}=\Omega^{2}E/(\omega^{2}-\Omega^{2})$;
the total elecric field acting upon the atomic nucleus is 
\begin{equation}
F=E+E_{i}=\frac{\omega^{2}}{\omega^{2}-\Omega^{2}}E_{0}\sin\omega t\:\:;\label{1}
\end{equation}
 since $\omega\ll\Omega$, we may use the approximation $F\simeq-(\omega^{2}/\Omega^{2})E\simeq-10^{-3}/Z^{2}$
(where $\omega=10^{15}s^{-1}$); we can see that the total field acting
upon the nucleus is appreciably reduced by the electron screening.
For $Z=50$ this reduction factor is $\simeq4\times10^{-7}$; the
maximum field $3\times10^{13}V/cm$ is reduced to $10^{7}V/cm$ ($\simeq10^{4}$
electrostatic units). It follows that we may limit ourselves here
to low fields acting upon the atomic nuclei. The case of strong fields
have been analyzed in Refs. \cite{key-8,key-9,key-18,key-22,key-23}.
At the same time, an induced electric field generated by the dipolar
eigenmodes occurs inside the atom, which oscillates with the higher
eigenfrequency $\Omega$.

If the field is low, the bound-state charge oscillates, emits higher-order
harmonics of electromagnetic radiation and tunneling may appear; in
this latter case, the charge accommodates itself in the field, in
a long time, which amounts to an adiabatically-introduced interaction;
this regime allows the usual, standard application of the tunneling
approach. As we shall see below, the threshold field which separates
the two regimes (low-field regime from high-field regime) can be obtained
from $\mid q\mid E_{0}/m\omega^{2}a=1$, where $a$ is a distance
of the order of the bound-state dimension (\emph{$a=10^{-13}cm$)}
(for protons, the threshold field is $E_{0}\simeq10^{5}V/cm$ ($10^{2}$
electrostatic units)).

Originally, the charge emission from bound states, like atom ionization,
has been treated by using adiabatic hypothesis, either by time-dependent
perturbation theory, or by imaginary-time tunneling, or other equivalent
approaches.\cite{key-24}-\cite{key-32} Quasi-classical tunneling
through the potential barrier generated by the field has been applied
in classical works to static fields and the hydrogen atom (in parabollic
coordinates).\cite{key-33}-\cite{key-35} For alpha-particle decay
or proton emission the situation is different. First, in spontaneous
decay, the alpha particle (and, in general, the ejected charge) is
preformed and, second, the tunneling through the Coulomb potential
barrier must be included.\cite{key-36}-\cite{key-39} We analyze
below the \textcolor{black}{spontaneous} charge emission, affected
by the presence of an adiabatically-introduced electromagnetic radiation,
in the presence of a Coulomb barrier; the problem may exhibit relevance
for studies of alpha-particle decay or proton emission. 

\textcolor{black}{The standard model of spontaneous alpha decay is
based on Bohr's concept of compound nuclei.\cite{key-40}}\textcolor{red}{{}
}\textcolor{black}{In an alpha-unstable nucleus the pre-formed alpha
particle acquires a kinetic energy and penetrates (tunnels through)
the Coulomb potential barrier. Consequently, the alpha-unstable nucleus
is in fact in a \textquotedbl{}metastable state\textquotedbl{}.} In
this simple model, the spontaneous alpha-particle decay and proton
emission proceed by tunneling through the Coulomb potential barrier,
as a result of many \textquotedbl{}attempts\textquotedbl{} the charge
makes to penetrate the barrier. The (high) frequency of this process
is of the order $1/t_{a}$, where $t_{a}$ corresponds, approximately,
to the energy level spacing; in atomic nuclei this spacing, for the
relevant energy levels, is of the order $\Delta\mathcal{E}=200keV$,
which gives $t_{a}\simeq3\times10^{-21}s$;\cite{key-40} also, the
broadening of the charge energy levels introduces an energy uncertainty
(we leave aside the so-called tunneling through the internal potential
barrier and the pre-formation factor of the alpha particle). The order
of magnitude of the energy of the charge is a few $MeV$, which ensures
a quasi-classical tunneling. The effect of the electromagnetic radiation
upon the initial preparation of the charge for tunneling may be neglected.
Similarly, we consider a sufficiently low electromagnetic radiation,
such that we may neglect its effects upon the mean-field potential.
We limit ourselves to the effect of the electromagnetic interaction
on the tunneling rate. 

Let us consider a charge $q>0$ with mass $m$ in the potential barrier
$V(\mathbf{r})$ in the presence of an electromagnetic radiation with
the vector potential $\mathbf{A}=\mathbf{A}_{0}\cos(\omega t-\mathbf{kr})$,
where $\mathbf{A}_{0}$ is the amplitude of the vector potential,
$\omega$ is the radiation frequency and $\mathbf{k}$ is the radiation
wavevector ($\omega=ck$); the electromagnetic field is transverse,
\emph{i.e.} $\mathbf{kA}=0$. Since the phase velocity of the non-relativistic
charge is much smaller than the speed of light $c$ in vacuum, we
may neglect the spatial phase $\mathbf{kr}$ in comparison with the
temporal phase $\omega t$; consequently, the vector potential may
be approximated by $\mathbf{A}\simeq\mathbf{A}_{0}\cos\omega t$.
This approximation amounts to neglecting the effects of the magnetic
field. It is assumed that this potential is introduced adiabatically.
The charge is immersed in the radiation field, such that we may start
with the standard non-relativistic hamiltonian 
\begin{equation}
H=\frac{1}{2m}\left(\mathbf{p}-\frac{q}{c}\mathbf{A}\right)^{2}+V(\mathbf{r})\:\:\:,\label{2}
\end{equation}
 where the momentum $\mathbf{p}$ includes the electromagnetic momentum
$q\mathbf{A}/c$ beside the mechanical momentum $m\mathbf{v}$, where
$\mathbf{v}$ is the velocity of the particle. We consider the Schrodinger
equation 
\begin{equation}
i\hbar\frac{\partial\psi}{\partial t}=H\psi\:\:;\label{3}
\end{equation}
since the interaction is time-dependent we need the time evolution
of the wavefunction. Consequently, in equation (\ref{3}) we perform
the well-known Kramers-Henneberger transform\cite{key-41}-\cite{key-44}
(with a vanishing electromagnetic interaction for $t\rightarrow-\infty$)
\begin{equation}
\begin{array}{c}
\psi=e^{iS}\varphi\:\:,\\
\\
S=\frac{q}{\hbar mc\omega}\mathbf{A}_{0}\mathbf{p}\sin\omega t-\frac{q^{2}A_{0}^{2}}{8\hbar mc^{2}\omega}(2\omega t+\sin2\omega t)\:\:;
\end{array}\label{4}
\end{equation}
 the Schrodinger equation becomes 
\begin{equation}
\begin{array}{c}
i\hbar\frac{\partial\varphi}{\partial t}=\frac{1}{2m}p^{2}\varphi+\widetilde{V}(\mathbf{r})\varphi\:\:,\\
\\
\widetilde{V}(\mathbf{r})=e^{-iS}V(\mathbf{r})e^{iS}=V(\mathbf{r}-q\mathbf{A}_{0}\sin\omega t/mc\omega)\:\:;
\end{array}\label{5}
\end{equation}
it is convenient to introduce the electric field $\mathbf{E}=\mathbf{E}_{0}\sin\omega t$,
$\mathbf{E}_{0}=\omega\mathbf{A}_{0}/c$; we get 
\begin{equation}
S=\frac{q}{\hbar m\omega^{2}}\mathbf{E}_{0}\mathbf{p}\sin\omega t-\frac{q^{2}A_{0}^{2}}{8\hbar mc^{2}\omega}(2\omega t+\sin2\omega t)\label{6}
\end{equation}
 and 
\begin{equation}
\widetilde{V}(\mathbf{r})=V(\mathbf{r}-q\mathbf{E}/m\omega^{2})\:\:.\label{7}
\end{equation}

We can see that for high-intensity fields the potential (including
the mean-field potential) is rapidly vanishing along the field direction.
Here we assume that the field intensity is low; specifically we assume
$qE_{0}/m\omega^{2}\ll a$, where $a$ is the dimension of the region
the charge moves in (the atomic nucleus); for protons this inequality
means $E_{0}\ll3\times10^{4}V/cm$ ($10^{2}$ electrostatic units),
as stated above. The preformed alpha particle (or emitted proton)
may tunnel through the potential barrier given by equation (\ref{7});
the \textquotedbl{}attempt\textquotedbl{} frequency to penetrate the
barrier and the energy uncertainty are practically not affected by
the low-intensity field. 

We adopt a model of nuclear decay by assuming a Coulumb potential
barrier $V(r)\simeq Zq^{2}/r$ (with the center-of-mass of the original
nucleus placed at the origin); in the absence of the field the tunneling
proceeds from $r_{1}=a$ to $r_{2}=Zq^{2}/\mathcal{E}_{r}$, where
$\mathcal{E}_{r}$ is the radial energy of the charge; it is convenient
to introduce the parameter $\xi=qE_{0}/m\omega^{2}a\ll1$, which includes
the effect of the field. In the presence of the field these limits
become
\begin{equation}
\widetilde{r}_{1}=\left|\mathbf{a}-q\mathbf{E}/m\omega^{2}\right|\label{8}
\end{equation}
and $\widetilde{r}_{2}=r_{2}$, where $\mathbf{a}=a\mathbf{r}/r$.
We expand $\widetilde{r}_{1}$ in powers of $\xi$ and get 
\begin{equation}
\widetilde{r}_{1}=a\left(1-\xi\sin\omega t\cdot\cos\theta+\frac{1}{2}\xi^{2}\sin^{2}\omega t\cdot\sin^{2}\theta\right)+...\:\:\:,\label{9}
\end{equation}
where $\theta$ is the angle the radius vector $\mathbf{r}$ makes
with the electric field $\mathbf{E}_{0}$.

To continue, we assume that the free charge attempting to penetrate
the potential barrier has momentum $\mathbf{p}_{n}$ and kinetic energy
$\mathcal{E}_{n}=p_{n}^{2}/2m$, where $n$ is a generic notation
for its state; we may leave aside the orbital motion and denote by
$\mathbf{p}_{rn}$ the radial momentum and by $\mathcal{E}_{rn}$
the radial energy. Let $\mathbf{p}_{r}$ and $\mathcal{E}_{r}=p_{r}^{2}/2m$
be the highest radial momentum and, respectively, the highest radial
energy; they correspond to the total momentum $\mathbf{p}$ and, respectiveley,
total energy $\mathcal{E}=p^{2}/2m$ (in general, a degeneration may
exist). This charge may tunnel through the potential barrier $V(r)$
from $\widetilde{r}_{1}$ to $\widetilde{r}_{2}$. The relevant factors
in the wavefunction $\psi$ given by equation (\ref{4}) are 
\begin{equation}
e^{\frac{iqE(t)}{\hbar m\omega^{2}}\cos\theta\cdot(p_{2}-p_{1})+\frac{i}{\hbar}\int_{\widetilde{r}_{1}}^{\widetilde{r}_{2}}dr\cdot p_{r}(r)}\:\:\:,\label{10}
\end{equation}
 where $p_{r}(r)=\sqrt{2m\left[\mathcal{E}-V(r)\right]}$, $p_{1,2}=p_{r}(\widetilde{r}_{1,2})=\sqrt{2m\left[\mathcal{E}-V(\widetilde{r}_{1,2})\right]}$;
it is easy to see that $p_{2}=0$. It follows that the tunneling probability
(transmission coefficient) is given by $w=e^{-\gamma}$, where
\begin{equation}
\begin{array}{c}
\gamma=-A\xi\sin\omega t\cdot\cos\theta+B\:\:,\\
\\
A=\frac{2a\left|p_{1}\right|}{\hbar}\:\:,\:\:\xi=\frac{qE_{0}}{m\omega^{2}a}\:\:,\:\:B=\frac{2}{\hbar}\int_{\widetilde{r}_{1}}^{\widetilde{r}_{2}}dr\left|p_{r}(r)\right|
\end{array}\label{11}
\end{equation}
and $\left|p_{1}\right|=\sqrt{2m\left[V(\widetilde{r}_{1})-\mathcal{E}\right]}$,
$\left|p_{r}(r)\right|=\sqrt{2m\left[V(r)-\mathcal{E}\right]}$ (the
condition $V(\widetilde{r}_{1})>\mathcal{E}$ ensures the existence
of the bound state). We expand the coefficient $A$ in powers of $\xi$
and take the average with respect to time; we get 
\begin{equation}
\gamma=-\frac{Zq^{2}}{2\hbar}\sqrt{\frac{2m}{Zq^{2}/a-\mathcal{E}}}\xi^{2}\cos^{2}\theta+B...\:\:;\label{12}
\end{equation}
the same procedure applied to the coefficient $B$ leads to 
\begin{equation}
\begin{array}{c}
B=\gamma_{0}-\frac{a\xi^{2}}{2\hbar}\sqrt{2m(Zq^{2}/a-\mathcal{E})}+\frac{a\xi^{2}}{2\hbar}\sqrt{\frac{2m}{Zq^{2}/a-\mathcal{E}}}(3Zq^{2}/2a-\mathcal{E})\cos^{2}\theta\:\:\:,\end{array}\label{13}
\end{equation}
where $\gamma_{0}$ corresponds to the absence of the radiation; finally,
we get 
\begin{equation}
\gamma=\gamma_{0}-\frac{a\xi^{2}}{2\hbar}\sqrt{2m(Zq^{2}/a-\mathcal{E})}\left[1-\frac{Zq^{2}/2a-\mathcal{E}}{Zq^{2}/a-\mathcal{E}}\cos^{2}\theta\right]\:\:.\label{14}
\end{equation}
 We can see that the effect of the radiation is to increase the rate
of charge emission by a factor proportional to the square of the electric
field ($\xi^{2}$) and to introduce a slight anisotropy. It is worth
noting that the radiation field contributes not only to the tunneling
factor, as expressed by the coefficient $B$, but it is present also
in the coefficient $A$, via the time-dependence of the wavefunction
provided by the Kramers-Henneberger transform. 

We can define a total disintegration probability 
\begin{equation}
w_{tot}\simeq\left\{ 1+\frac{a\xi^{2}}{2\hbar}\sqrt{2m(Zq^{2}/a-\mathcal{E})}\left[1-\frac{Zq^{2}/2a-\mathcal{E}}{3(Zq^{2}/a-\mathcal{E})}\right]\right\} w_{tot}^{0}\label{15}
\end{equation}
by integrating over angle $\theta$, where $w{}_{tot}^{0}=e^{-\gamma_{0}}$.
The disintegration rate per unit time is $(1/\tau)w_{tot}$, where
$\tau$ is related to the time $t_{a}$ estimated above and the time
introduced by the energy uncertainty.\cite{key-40} 

The exponent $\gamma_{0}$, corresponding to the absence of the radiation,
is 
\begin{equation}
\gamma_{0}=\frac{Zq^{2}}{\hbar}\sqrt{2m/\mathcal{E}}\left(\arccos\sqrt{\mathcal{E}a/Zq^{2}}-\sqrt{\mathcal{E}a/Zq^{2}}\sqrt{1-\mathcal{E}a/Zq^{2}}\right)\:\:;\label{16}
\end{equation}
since $Zq^{2}/a\gg\mathcal{E}$ (for protons $q^{2}/a=2.5MeV$) ,
we may use the approximate formulae 
\begin{equation}
\gamma_{0}\simeq\frac{\pi Zq^{2}}{2\hbar}\sqrt{2m/\mathcal{E}}\label{17}
\end{equation}
 and 
\begin{equation}
w_{tot}\simeq\left(1+\frac{5a\xi^{2}}{12\hbar}\sqrt{2mZq^{2}/a}\right)w_{tot}^{0}\:\:.\label{18}
\end{equation}
As it is well know the interplay between the very large values of
$1/\tau$ and the very small values of $e^{-\gamma_{0}}$, makes the
disintegration rate to be very sensitive to the energy values, and
to vary over a wide range.\cite{key-40} The result can be cast in
the form of the Geiger-Nuttall law, which, in the absence of the radiation,
can be written as $\ln(w_{tot}^{0}/\tau)=-a_{0}Z/\sqrt{\mathcal{E}}+b_{0}$,
$a_{0}$ and $b_{0}$ being well-known constants;\cite{key-40} the
only effect of the radiation is to modify the constant $b_{0}$ into
$b=b_{0}+(5a\xi^{2}/12\hbar)\sqrt{2mZq^{2}/a}$. The correction to
$b_{0}$ can also be written as $(5\xi^{2}/12)[(Zq^{2}/a)/(\hbar^{2}/2ma^{2})]^{1/2}$
for $\xi\ll1$. \textcolor{black}{The maximum value of this correction
is of the order of the unity; it follows that the decay rate is enhanced
by the radiation by a factor of the order }$\xi^{2}\ll1$.\textcolor{black}{{} }

After the emission of the charge, the mean-field potential suffers
a reconfiguration (re-arrangement) process and the potential $V(\mathbf{r})$
is modified; this is the well-known process of \textquotedbl{}core
shake-up\textquotedbl{} (or \textquotedbl{}core excitation\textquotedbl{});
a new bound state is formed and a new transformation process may begin
for the modified potential $V(\mathbf{r})$. The tunneling probability
$w$ given above is a transmission coefficient (we can check that
$w<1$); with probability $1-w$ the charge is reflected from the
potential barrier; in these conditions the bound state is \textquotedbl{}shaken-up\textquotedbl{}
and the charge resumes its motion, or its preformation process, untill
it tunnels, or is rescattered back to the core; the latter is the
well-known recollision process.\cite{key-8},\cite{key-45}-\cite{key-49}

The case of a static field requires a special discussion. Within the
present formalism a static electric field $\mathbf{E}$ can be obtained
from a vector potential $\mathbf{A}=-c\mathbf{E}t$; the position
vector in the mean-field potential is shifted to $\mathbf{r}\rightarrow\mathbf{r}+\boldsymbol{\zeta}$,
where $\boldsymbol{\zeta}=q\mathbf{E}t^{2}/2m$; the special discussion
is necessary because the parameter $\boldsymbol{\zeta}$ is unbounded
in time. The distance $a$ is covered in time $t_{0}=\sqrt{2ma/qE}$;
for proton, $a$ is of the order $a=10^{-13}cm$ and the threshold
field is $E=E_{0}=3\times10^{4}V/cm$ ($10^{2}$ electrostatic units)
given above; we get $t_{0}\simeq10^{-15}s$. This is a very long duration,
in comparison with the relevant nuclear times, in particular the attempt
time $\tau$ ($t_{a}\simeq10^{-21}s$ estimated above). In general,
the condition of adiabatic interaction reads $t_{0}\ll\hbar/\Delta\mathcal{E}$,
where $\Delta\mathcal{E}$ is the mean separation of the energy levels;
it implies $qEa\ll(\Delta\mathcal{E})^{2}/(\hbar^{2}/ma^{2})$, which
allows for high static fields. In these conditions the protons accommodate
themselves to the electric field, which is absorbed into slightly
modified energy levels; this change, which can be estimated by perturbation
theory, is irrelevant for our discussion, since the field strength
is small. However, it has an important consequence in that the electric
field, once taken in the energy levels, is not available anymore for
the Kramers-Henneberger transform given by equation (\ref{4}); therefore,
the present time-dependent formalism cannot be applied. Instead of
using the hamiltonian given by equation (\ref{2}), we start with
the (equivalent) dipole hamiltonian which includes the interaction
term $-q\mathbf{Er}$. Consequently, the potential barrier $V(r)\simeq Zq^{2}/r$
is changed into 
\begin{equation}
V(\mathbf{r})=\frac{Zq^{2}}{r}-q\mathbf{Er}=\frac{Zq^{2}}{r}\left(1-\frac{Er^{2}}{Zq}\cos\theta\right)\:\:.\label{19}
\end{equation}
We compute the tunneling rate by using this potential barrier. In
view of the small value of the correction parameter proportional to
$E$ in equation (\ref{19}), we may expand the momentum $p_{r}=\sqrt{2m\left[\mathcal{E}-V(\mathbf{r})\right]}$
in powers of this parameter and replace the powers of $r^{2}$ by
their mean values over the tunneling range from $r_{1}=a$ to $r_{2}=Zq^{2}/\mathcal{E}$;
since $r_{2}\gg a$, we get the small parameter $\alpha=Er_{2}^{2}/Zq=Eqr_{2}/\mathcal{E}\ll1$
in equation (\ref{19}). For $Z=100$ and $\mathcal{E}=1MeV$ this
parameter is $\alpha=10^{-4}E$, which is much smaller than unity
for any usual static field. Integrating over angles and assuming $\alpha\gamma_{0}\ll1$,
where $\gamma_{0}$ is given by equation (\ref{17}), we get finally
\begin{equation}
w_{tot}\simeq\left(1+\frac{\alpha^{2}\gamma_{0}^{2}}{108}\right)w_{tot}^{0}\:\:.\label{20}
\end{equation}
We can see that the correction brought by a static electric field
to the decay rate is extremely small, as expected. 

Finally, it is worth discussing the case of intermediate fields, \emph{i.e.}
field strengths which satisfy the inequality $qE_{0}/m\omega^{2}>a$
($\xi>1$) (in our case, fields from $3\times10^{4}V/cm$ to $10^{7}V/cm$).\cite{key-50}
In this case the adiabatic hypothesis cannot be used anymore, and
the initial conditions for introducing the interaction are important.
The corresponding Kramers-Henneberger transform diminishes appreciably
the potential barrier and the charge is set free in a short time,
which is the reciprocal of the decay rate; this rate may exhibit oscillations
as a function of the field strength. 

In conclusion, we may say that in low-intensity electromagnetic radiation
the bound-states charges accommodate themselves in the field, which
amounts to an adiabatically-introduced interaction, as it is well
known. In these conditions, besides oscillating and emitting higher
harmonics, the charge may tunnel out from the bound state. This is
the standard ionization process, which was widely investigated for
atom ionization. In spontaneous alpha decay or proton emission the
situation is different, because of the preformation stage and the
tunneling through the Coulomb potential barrier. We have analyzed
above the disintegration rate for the charge emission from atomic
nuclei in the case of the adiabatic introduction of electromagnetic
interaction, with application to nuclear alpha-particle decay and
proton emission. Under these circumstances, it has been shown in this
paper that the tunneling rate (through Coulomb potential) is slightly
enhanced by the presence of the radiation, by corrections whose leading
contributions are of second-order in the electric field, with a slight
anisotropy. Similar results are presented in this paper for static
fields. 

\textbf{Acknowledgements.} The author is indebted to S. Misicu and
the members of the Laboratory of Theoretical Physics at Magurele-Bucharest
for many fruitful discussions. This work has been supported by the
Scientific Research Agency of the Romanian Government through Grants
04-ELI / 2016 (Program 5/5.1/ELI-RO), PN 16 42 01 01 / 2016 and PN
(ELI) 16 42 01 05 / 2016.

\end{document}